# DETERMINATION OF SPIN AND OTHER QUANTUM CHARACTERISTICS OF NEUTRAL BOSON WITH MASS AROUND 126 GeV DISCOVERED IN CMS AND ATLAS EXPERIMENTS AND ITS IDENTIFICATION

**Vali A. Huseynov*[1,2]**

[1] Department of Theoretical Physics, Baku State University,
Z. Khalilov 23, AZ 1148, Baku, Azerbaijan;
[2] Department of General and Theoretical Physics, Nakhchivan State University,
University Campus, AZ 7012, Nakhchivan, Azerbaijan
E-mail: vgusseinov@yahoo.com

*In my previous journal papers my surname was Guseinov instead of Huseynov.

**Abstract**

We investigate the single neutral bosons (NB) with the spins 0, 1 and 2 decaying via the $W^-W^+$-channel in an external magnetic field (EMF) and discuss the questions connected with the search of the Standard Model (SM) scalar Higgs boson (HB) in the CMS and ATLAS experiments at the LHC. It is shown that a single neutral scalar boson with the mass around $126\,GeV$ can not decay into the two on-shell $W^-W^+$-bosons in an EMF. The impossibility of decay of a single neutral scalar boson in the mass range below $160.770\,GeV$ into the two on-shell $W^-W^+$-bosons in an EMF due to the energy and spin projection conservation laws and the possible decay of a single NB with the spin $J=2$ and the spin projection $J_z=+2$ in that region enable us to come to the conclusion that the single NB with the mass $125.3\,GeV/126\,GeV$ discovered in the CMS and ATLAS experiments is neither the SM HB nor a scalar boson at all. The NB with the mass $125.3\,GeV/126\,GeV$ discovered in the CMS and ATLAS experiments is a new NB with the spin $J=2$ and the spin projection $J_z=+2$ that is not included in the SM. Both the $P$-parity and charge conjugation $C$ of this new particle are +1. So, the newly discovered NB is a neutral tensor boson that is characterized by $J^{PC}=2^{++}$ quantum numbers under parity $P$ and charge conjugation $C$ and $CP=+1$. The other quantum characteristics of this NB are as follows: the orbital quantum number $L=0$, the weak isospin $T=2$, the third component of the weak isospin $T_z=0$ and the weak hypercharge $Y_W=0$. The quantum state of the discovered NB is characterized by the $1^5S_2$ spectral term. At the same time the observed NB with the mass around $126\,GeV$ can be considered as a fundamental particle that is a carrier (quantum) of the interactions between bosons, at least between electroweak bosons. To clarify this situation more and new experimental data are required.

## 1. INTRODUCTION

The Standard Model of elementary particles is based on the electroweak symmetry group $SU(2)_L \times U(1)_Y$ [1-3] and strong $SU(3)_C$ group of QCD [4-7] and predicted the existence of $W^{\pm}$- and $Z$-bosons that were experimentally confirmed by the U1 and U2 Collaborations [8-12].

In the Standard Model the spontaneous symmetry breaking Higgs mechanism [13-18] enables the fermions and gauge bosons to obtain a mass. When global symmetry is spontaneously broken, the massless Goldstone bosons (spin 0) appear [19, 20]. The $W^{\pm}$- and $Z$-bosons "eat" these three Goldstone bosons to form their longitudinal components and obtain their masses. The fourth neutral component of the complex scalar doublet corresponds to the scalar Higgs boson. The Higgs boson gives the mass for all the fermions and itself. The Higgs boson predicted by the Standard Model is characterized by $J^{PC} = 0^{++}$ quantum numbers under the intrinsic parity $P$ and the charge conjugation $C$. Since the $U(1)$ symmetry is unbroken, the photon remains massless. The mass of the Higgs boson is not predicted by the Standard Model. One of the most important problems of modern particle physics is to determine whether the electroweak symmetry breaking Higgs mechanism is realized in nature. How much mass has the Higgs boson if the only one complex scalar isodoublet allows mass generation for all massive particles of the Standard Model and the Higgs boson itself due to electroweak symmetry breaking?

The following limit $m_H < 158\, GeV$ at 95% confidence level (CL) [21, 22] is derived for the Standard Model Higgs mass on the basis of the global fits to precision electroweak results. The $e^+e^-$-collision experiments at the LEP have led to a lower mass bound of $m_H > 114.4\, GeV$ for the Standard Model Higgs boson [23]. The results of the Tevatron $p\bar{p}$-collision experiments have excluded at 95% CL the regions $147\, GeV < m_H < 179\, GeV$ and $100\, GeV < m_H < 106\, GeV$ [24-27]. On the 4 July 2012 the ATLAS and CMS Collaborations reported on the discovery of a new "higgs-like" neutral boson with a measured mass of $126 \pm 0.4\,(stat.) \pm 0.4\,(sys.)\, GeV$ [28] and $125.3 \pm 0.4\,(stat.) \pm 0.5\,(sys.)\, GeV$ [29], respectively.

The Standard Model Higgs boson in the mass ranges $112.9\, GeV - 115.5\, GeV$,

$131\,GeV - 238\,GeV$ and $251\,GeV - 466\,GeV$ was excluded by the ATLAS Collaboration [30]. The mass range $127\,GeV - 600\,GeV$ at 95% CL and the mass range $129\,GeV - 525\,GeV$ at 99% CL [31] for the Standard Model Higgs boson were excluded by the CMS Collaboration. The recent analysis of combined search for the Standard Model Higgs boson in $pp$-collisions at $\sqrt{s} = 7\,TeV$ with the ATLAS detector excluded the mass ranges of $111.4\,GeV$ to $116.6\,GeV$, $119.4\,GeV$ to $122.1\,GeV$ and $129.2\,GeV$ to $541\,GeV$ at the 95% CL [32].

The analysis show that the Standard Model Higgs boson mass range between $111.4\,GeV$ and $600\,GeV$ at the 95% CL has been excluded except $116.6\,GeV - 119.4\,GeV$, $122.1\,GeV - 127\,GeV$.

The four main production channels for the single Standard Model scalar Higgs boson at the LHC are the gluon-gluon fusion channel ($gg \to H$), the vector boson fusion channel ($qq' \to qq'H$), the Higgs-strahlung channel ($qq' \to WH, ZH$) and the channel of the associated production with a $t\bar{t}$ pair ($q\bar{q}/gg \to t\bar{t}H$) [33-35].

The neutral boson with the mass around $126\,GeV$ discovered by the ATLAS and CMS Collaborations is instable and decays on different channels. This neutral boson was discovered at the LHC while studying the decay channels $H \to \gamma\gamma$, $H \to e^+e^-e^+e^-$, $H \to e^+e^-\mu^+\mu^-$, $H \to \mu^+\mu^-\mu^+\mu^-$ [28, 29]. The decay of this boson into the leptons proceeds at two stages: firstly a new discovered neutral boson decays into the two known heavy neutral particle, i.e. into the two $Z$-bosons. One of these $Z$-bosons is a virtual one. Then each of the $Z$-bosons decays into $e^+e^-$-pair, or $\mu^+\mu^-$-pair. It can be written as $H \to ZZ^* \to 4l$, where $Z^*$ is the virtual boson and $l$ is one of the leptons $e^\pm, \mu^\pm$. According to the report of the both collaborations some excess of events connected with the decay $H \to WW^* \to l\nu l\nu$ are observed where $W$ is a charged vector boson and $\nu$ is an electron (or muon) neutrino [28, 29] (see also Ref. [35]).

In this work we investigate the decay of single neutral bosons with the spins 0,

1 and 2 into the two $W^{\pm}$-bosons in an external magnetic field. The interest to the presented investigation is motivated by the following experimental data and facts on the newly discovered neutral boson with the mass around $126\,GeV$ collected in the ATLAS and CMS experiments at the LHC and also the naturally arising questions connected with the spin, the intrinsic parity, the charge conjugation and the other important quantum characteristics of this neutral boson.

1) According to the Standard Model measurable are $b\bar{b}$, $\tau\bar{\tau}$, $WW^*$, $ZZ^*$ and $\gamma\gamma$ decay channels of the Higgs boson. At the same time the decay mode $H \to Z\gamma$ is also possible. In more than half of the all cases ($\geq 50\%$) the Standard Model Higgs boson is to decay into the $b\bar{b}$-pair. This decay mode has maximum branching ratio for the Higgs mass around $125\,GeV$. The probability of the $H \to \tau^+\tau^-$ decay mode is 6% for $m_\tau = 1.8\,GeV$. The Standard Model Higgs boson is to decay on the channel $H \to ZZ^* \to 4l$ with the probability $0.013\%$. The probability of the decay channel $H \to \gamma\gamma$ is $0.23\%$. The decay probability of the Standard Model Higgs boson on the channel $H \to Z\gamma$ is $0.15\%$ [35]. However, the current experimental data on the Higgs boson signals collected from the ATLAS and CMS experiments show some deviations from the Standard Model expectation. One of the most significant deviations from the Standard Model expectation is connected with the enhancement in the $H \to \gamma\gamma$ decay rate. As we have just noted that the $H \to \gamma\gamma$ decay is to have a small relative probability in the case of the existence of one complex scalar isodoublet. It was indicated in [36] that this decay can become essential and in principle, even dominant if there are more than one Higgs boson (see also Ref. [37-38]). This is one of the numerous possible versions. What is the real reason of these significant deviations from the Standard Model expectation?

2) Although the decay probability of the Standard Model Higgs boson on the channel $H \to b\bar{b}$ or $H \to \tau^+\tau^-$ is much more than the probabilities in other channels, some deficits in the $b\bar{b}$ and $\tau^+\tau^-$ channels were observed in the experiments carried out by the ATLAS and CMS collaborations. What is the reason of these deficits? Can, in principle, the newly discovered neutral boson with the mass $125.3\ GeV/126\ GeV$ (hereafter we call it $Y$-boson) decay on the $b\bar{b}$ and $\tau^+\tau^-$ channel?

3) As we have just noted that the neutral $Y$-boson with the mass $125.3\ GeV/126\ GeV$ was discovered in the $Y \to \gamma\gamma$ reaction which is one of the main decay channels. The observation of this reaction channel and the Landau-Yang theorem tell that the spin of the observed new particle can not be 1 [39, 40]. Because the particle with the spin 1 can not decay into the two photons which have two polarizations ($\pm 1$). The ATLAS collaboration reports: "The observation in the diphoton channel disfavours the spin 1 hypothesis" [28]. A single neutral particle with spin 0, or 2 or even higher than spin 2 can decay into the two photons [41]. So, there are other possibilities left: either the new discovered particle is a neutral boson with the spin 0 or a neutral boson with the spin 2 or even higher than spin 2 [35]. One of the main missions of the current experimental and theoretical investigations on the newly discovered neutral boson with the mass $125.3\ GeV/126\ GeV$ is to determine the spin, the intrinsic parity, the charge conjugation and the other important quantum characteristics of this neutral boson, to clarify whether this is the Standard Model Higgs boson or it is another unknown particle and to identify this particle.

4) According to the results of the analysis of the experimental data the production of the Standard Model Higgs boson is, in principle, possible or is not excluded in the mass ranges $116.6 GeV - 119.4\ GeV$ and $122.1 GeV - 127\ GeV$. The following questions arise in this situation. Can the Higgs boson or any other single scalar boson with the mass around $126\ GeV$ decay into the two on-shell $W^\pm$-bosons in an external magnetic field? In what mass range ought to we search for the Standard Model Higgs boson if it really exists and decays into the two $W^\pm$-bosons?

The main purpose of our work presented here is to determine all possible polarization states of the single neutral bosons with the spins $J = 0, 1, 2$ decayed into the two $W^\pm$-bosons, to show the impossibility of decay of the "higgs-like" boson with the mass around $126\ GeV$ discovered in the CMS and ATLAS experiments at the LHC, into the two $W^\pm$-bosons, to determine the spin, the spin projection onto the $Oz$-axis, the charge conjugation, the spectral term, the intrinsic parity, the weak isospin, the third component of the weak isospin, the weak hypercharge of this neutral "higgs-like" boson, to prove that the discovered neutral single boson with the mass around $126\,GeV$ is not a scalar boson, to identify this neutral boson and to clarify the mass range where we ought to search for the Standard Model Higgs boson if it really exists and decays into the two $W^{\pm}$-bosons.

The other important purpose of the presented work is to clarify the problems connected with the following questions. Can the Higgs boson or any other single scalar boson with the mass around $126\,GeV$ be created via the charged vector boson (on-shell $W^{\pm}$-boson ) fusion channel or decay into the two on-shell $W^{\pm}$-bosons in an external magnetic field? Can the Higgs boson or any other single scalar boson with the mass around $126\,GeV$ be created from the vacuum at the expense of the virtual (off-shell) $W^{\pm}$-bosons in an external electric field?

## 2. $W$-BOSON IN EXTERNAL MAGNETIC FIELD. DECAY OF SINGLE NEUTRAL BOSONS WITH SPINS $J = 0, 1, 2$ VIA $W^{-}W^{+}$- CHANNEL IN EXTERNAL MAGNETIC FIELD

### 2.1. Energy spectrum of $W$-boson in external constant homogenous magnetic field

If a single neutral boson $Y$ is produced via a virtual (off-shell) $W$-boson fusion channel (e.g., a Higgs-like boson in the LHC experiments), it can also be produced via a real (on-shell) $W$-boson pair fusion channel in a free case or in an external magnetic field. So, it is possible to investigate the production of a single neutral boson $Y$ at the expense of the two real (on-shell) $W^{\pm}$-bosons in the reaction $W^{-}W^{+} \rightarrow Y$ in an external magnetic field. In this indicated case the two real (on-shell) $W^{-}$- and $W^{+}$-bosons are placed in an external constant homogenous magnetic field. Production of the single neutral boson with the mass around $126\,GeV$ (we consider that the produced single neutral boson energy is $E_Y = m_Y$, i.e. it is at rest) is possible at the expense of the reaction $W^-W^+ \to Y$ if the two real (on-shell) $W^-$- and $W^+$-bosons with the definite spin polarization states are placed in an external constant homogenous magnetic field with the given strength [42]. In [42] we investigated the production of single neutral bosons with spin 0 and 2 in the reaction $W^-W^+ \to Y$.

On the other hand a neutral "higgs-like" boson decays via the $H \to WW^* \to l\nu l\nu$ channel [43]. One of these bosons ($W$) is on-shell, the other boson ($W^*$) is off-shell. According to the energy conservation law the decay of the "higgs-like" boson with the mass around $126\,GeV$ into the two on-shell $W$-bosons is impossible. Therefore the "higgs-like" boson with the mass around $126\,GeV$ decays into one on-shell $W$-boson and one off-shell $W^*$-boson. However, if we place the "higgs-like" boson with the mass around $126\,GeV$ (we call it $Y$-boson) in an external magnetic field and consider that the decayed single neutral boson is at rest, i.e. $E_Y = m_Y$, its decay into the two on-shell $W$-bosons is possible, i.e. this decay is allowed by the energy conservation law. We will show it a bit later.

The below obtained results and calculations are true for both the $W^-W^+ \to Y$ reaction and $Y \to W^-W^+$ reaction. Therefore we investigate one of these reactions in a constant homogenous magnetic field. It should be noted that in [42] we investigated the production of single neutral bosons with the spins $0$ and $2$ on the basis of the reaction $W^-W^+ \to Y$ in an external magnetic field and determined the spin ($J=2$) and the spin projection ($J_z = +2$) of the single neutral boson with the mass around $126\,GeV$.

So, here we start to investigate the decay of the single neutral bosons with the spins $0, 1, 2$ including the "higgs-like" boson with the mass around $126\,GeV$ into the two on-shell $W$-bosons in an external constant homogenous magnetic field

$$Y \to W^- + W^+. \qquad (1)$$

We place a single neutral boson $Y$ in a magnetic field and we consider this boson in the rest frame, i.e. $E_Y = m_Y$. In a magnetic field a new reaction channel is opened: a single neutral boson $Y$ in the rest frame in a constant homogenous magnetic field can decay into the two real (on-shell) $W^-W^+$-pair.

Firstly, let us consider the energy spectrum of a $W$-boson in an external constant magnetic field [44-48] (see also Ref. [48])

$$E_n^2 = p_z^2 + (2n+1)e|\vec{B}| - 2e\vec{B}\cdot\vec{s} + m_W^2, \qquad (2)$$

where we assumed magnetic field vector $\vec{B}$ to be directed along the $Oz$ axis, $|\vec{B}| = B$ is the strength of an external magnetic field, $m_W$ is the $W$-boson mass ($m_W \cong 80.385\ GeV$ [49]), $p_z$ and $\vec{s}$ are the third component of a momentum and the spin of a $W$-boson, respectively, $n$ denotes the number of the Landau level ($n = 0,1,2,...$). The projection of the spin of a $W$-boson which is a massive vector boson onto the $Oz$ axis has three possible values: $s_z = 0, \pm 1$.

Hereafter we will consider the case

$$n = 0,\ p_z = 0. \qquad (3)$$

When the spin of a $W$-boson is oriented along the magnetic field direction ($\theta = \hat{\vec{B}\vec{s}} = 0$), i.e. when $s_z = 1$, the energy spectrum of a $W$-boson is determined as

$$E^2 = m_W^2 - eB. \qquad (4)$$

When the spin of a $W$-boson is orthogonal to the magnetic field direction ($\theta = 0$), i.e. when $s_z = 0$, the energy spectrum is determined as

$$E^2 = m_W^2 + eB. \qquad (5)$$

If the spin of a $W$-boson is oriented opposed to the magnetic field direction ($\theta = \pi$), i.e. when $s_z = -1$, the energy spectrum is determined as follows

$$E^2 = m_W^2 + 3eB. \qquad (6)$$

## 2.2. Possible polarization states of single neutral bosons with spin $J = 0, 1, 2$ decayed via $W^-W^+$-channel

A $W^\pm$-bosons have the following three polarization states depending on the projection of its spin onto the $Oz$ axis ($s_z = 0, \pm 1$):

$$\left|W^{\pm}(s_{\pm}=1,\ s_{\pm z}=+1)\right\rangle = \left|1,+1\right\rangle, \tag{7}$$

$$\left|W^{\pm}(s_{\pm}=1,\ s_{\pm z}=0)\right\rangle = \left|1,0\right\rangle, \tag{8}$$

$$\left|W^{\pm}(s_{\pm}=1,\ s_{\pm z}=-1)\right\rangle = \left|1,-1\right\rangle, \tag{9}$$

where we have denoted the spin of the $W^-(W^+)$-boson with $s_-=1(s_+=1)$ and the projection of the spin of the $W^-(W^+)$-boson onto the $Oz$ axis with $s_{-z}=0,\pm1(s_{+z}=0,\pm1)$.

Here we develop the idea of decay of the single neutral bosons $Y$ with the spins $J=0,1,2$ into a real (on-shell) $W^-$- and $W^+$-bosons. We consider the decayed single neutral bosons $Y$ have various polarization states depending on the spin projection onto the $Oz$ axis ($J_z$). The polarization states of the single neutral bosons decayed into the two on-shell $W$-bosons can really have the spin equal to 2, 1, 0. Using the polarization states (7)-(9), the addition rule of spins and the Clebsh-Gordan coefficients [41, 49] we can present the following possible quantum (polarization) states $|J,J_z\rangle$ of the decayed single neutral bosons $Y$ in the various $Y \to W^- + W^+$ reactions depending on the polarization states of $W^{\pm}$-bosons:

$$|2,+2\rangle = |1,+1;1,+1\rangle, \tag{10}$$

$$|2,+1\rangle = \frac{1}{\sqrt{2}}(|1,+1;1,0\rangle + |1,0;1,+1\rangle), \tag{11}$$

$$|2,0\rangle = \frac{1}{\sqrt{6}}(2|1,0;1,0\rangle + |1,+1;1,-1\rangle + |1,-1;1,+1\rangle), \tag{12}$$

$$|2,-1\rangle = \frac{1}{\sqrt{2}}(|1,0;1,-1\rangle + |1,-1;1,0\rangle), \tag{13}$$

$$|2,-2\rangle = |1,-1;1,-1\rangle, \tag{14}$$

$$|1,+1\rangle = \frac{1}{\sqrt{2}}(|1,+1;1,0\rangle - |1,0;1,+1\rangle), \tag{15}$$

$$|1,0\rangle = \frac{1}{\sqrt{2}}(|1,+1;1,-1\rangle - |1,-1;1,+1\rangle), \tag{16}$$

$$|1,-1\rangle = \frac{1}{\sqrt{2}}(|1,0;1,-1\rangle - |1,-1;1,0\rangle), \tag{17}$$

$$|0,0\rangle = \frac{1}{\sqrt{3}}(|1,+1;1,-1\rangle + |1,-1;1,+1\rangle - |1,0;1,0\rangle). \tag{18}$$

The polarization states (10)-(14) have the spin 2. Therefore these polarization states belong to the class of tensor particles. Let us consider the reactions of decay of the tensor polarization states into the $W^+$- and $W^-$-bosons.

### 2.3. Polarization state $|2,+2\rangle$

The polarization state $|2,+2\rangle$ decays according to the reaction

$$Y(J=2, J_z=+2) = W^-(s_-=1, s_{-z}=+1) + W^+(s_+=1, s_{+z}=+1). \tag{19}$$

When the $W^-$- and $W^+$-bosons having the spin projections $s_{-z}=+1$, $s_{+z}=+1$, respectively, occupy the quantum state with $n=n'=0$ and $p_z = p'_z = 0$, the corresponding energy conservation law

$$m_Y = \sqrt{m_W^2 - eB} + \sqrt{m_W^2 - eB} \tag{20}$$

for the reaction channel (19) is written on the basis of the expression (2). The equality (20) shows that the strength of an external magnetic field changes in the following range

$$0 \leq B \leq \frac{m_W^2}{e}. \tag{21}$$

On the basis of the energy conservation law (20) and the inequality (21) we derive the following mass range

$$0 \leq m_Y \leq 2m_W \tag{22}$$

or

$$0 \leq m_Y \leq 160.770 \ GeV. \tag{23}$$

for the production of a single neutral boson with the spin $J=2$ and the spin projection $J_z=+2$ via the reaction channel (19).

The mass 125.3 $GeV$/126 $GeV$ of a single neutral boson discovered in the CMS

and ATLAS experiments is in the mass range (22) or (23). The mass range (22) or (23) corresponds to the decay of a single neutral boson with the spin $J = 2$ and the spin projection $J_z = +2$ provided that a single neutral boson with the mass $125.3\ GeV / 126\ GeV$ decays into the known $W^-$-boson and $W^+$-boson.

### 2.4. Polarization state $|2, +1\rangle$

The polarization state $|2, +1\rangle$ decays according to the reaction

$$Y(J = 2, J_z = +1) = W^-(s_- = 1, s_{-z} = +1) + W^+(s_+ = 1, s_{+z} = 0), \quad (24)$$

$$Y(J = 2, J_z = +1) = W^-(s_- = 1, s_{-z} = 0) + W^+(s_+ = 1, s_{+z} = +1). \quad (25)$$

When both of the $W^\pm$-bosons occupy the quantum state with $n = n' = 0$ and $p_z = p'_z = 0$, the corresponding energy conservation law

$$m_Y = \sqrt{m_W^2 - eB} + \sqrt{m_W^2 + eB} \quad (26)$$

for the reaction channels (24) and (25) is written on the basis of the expression (2). On the basis of the inequality (21) and the equality (26) we derive the following mass range

$$\sqrt{2}m_W \leq m_Y \leq 2m_W. \quad (27)$$

or

$$113.682\ GeV \leq m_Y \leq 160.770\ GeV. \quad (28)$$

So, the mass of the single neutral boson with the spin $J = 2$ and the spin projection $J_z = +1$ which corresponds to the polarization state $|2, +1\rangle$ is in the mass range (27) or (28).

### 2.5. Polarization state $|2, 0\rangle$

The contributions to the polarization state $|2, 0\rangle$ are given by the reactions

$$Y(J = 2, J_z = 0) = W^-(s_- = 1, s_{-z} = +1) + W^+(s_+ = 1, s_{+z} = -1), \quad (29)$$

$$Y(J = 2, J_z = 0) = W^-(s_- = 1, s_{-z} = -1) + W^+(s_+ = 1, s_{+z} = +1), \quad (30)$$

$$Y(J = 2, J_z = 0) = W^-(s_- = 1, s_{-z} = 0) + W^+(s_+ = 1, s_{+z} = 0). \quad (31)$$

When both of the $W^{\pm}$-bosons occupy the quantum state with $n = n' = 0$ and $p_z = p'_z = 0$, the corresponding energy conservation law is written as

$$m_Y = \sqrt{m_W^2 - eB} + \sqrt{m_W^2 + 3eB} \qquad (32)$$

for the reaction channels (29) and (30). The equality (32) shows that the strength of an external magnetic field changes in the range determined with the inequality (21). On the basis of the inequality (21) and the equality (32) we derive the following mass range

$$2m_W \leq m_Y \leq \frac{4\sqrt{3}}{3} m_W, \qquad (33)$$

or

$$160.770 \, GeV \leq m_Y \leq 185.641 \, GeV \qquad (34)$$

for the single neutral boson with the spin $J = 2$ and the spin projection $J_z = 0$ which corresponds to the polarization state $|2, 0\rangle$. When the single neutral boson with the spin $J = 2$ and the spin projection $J_z = 0$ decays into the two on-shell $W$-bosons in a constant homogenous magnetic field and one of the $W$-boson has the spin projection $s_{\mp z} = \pm 1$, the other $W$-boson has the spin projection $s_{\pm z} = \mp 1$, the maximal value of the mass of this single neutral boson corresponds to the magnetic field strength

$$B = \frac{2}{3} B_{0W} = \frac{2}{3} \frac{m_W^2}{e}. \qquad (35)$$

Thus, a single neutral tensor boson with the mass less than $2m_W \cong 160.770 \, GeV$ can not decay into the two on-shell $W$-bosons in a constant homogenous magnetic field when one of the $W$-bosons occupies the state with the spin projection $s_{\mp z} = +1$ and the other $W$-boson occupies the state with the spin projection $s_{\pm z} = -1$.

Now let us consider the single neutral tensor boson decay channel (31) in an external magnetic field when both of the $W^{\pm}$- bosons occupy the state with the spin projection $s_{\pm z} = 0$. On the basis of the energy and spin projection conservation laws we can write the following expression for the considered reaction channel

$$m_Y = \sqrt{m_W^2 + eB} + \sqrt{m_W^2 + eB}. \qquad (36)$$

The minimal value of the mass of a single neutral tensor boson $|2, 0\rangle$ produced via the channel (31) is

$$m_Y = 2m_W \cong 160.770 \, GeV \qquad (37)$$

and corresponds to the magnetic field strength of $B = 0$. At the critical value of the magnetic field strength $B = B_{0W} = m_W^2/e$ the mass of the single neutral tensor boson produced via the channel (31) in an external magnetic field is

$$m_Y(B = B_{0W}) = 2\sqrt{2}m_W \cong 227.363 \, GeV. \qquad (38)$$

Apart from the $W$-boson state $s_{\pm z} = +1$ and the expression (32) the $W$-boson state with $s_{\pm z} = 0$ and the equality (36) tells us that there is no restriction for the value of the strength of an external magnetic field when we have dealing with the $W$-bosons with the spin projections $s_{\pm z} = 0$. It means that, in principle, $B$ can change in the range

$$0 \leq B < \infty. \qquad (39)$$

So, the mass of the single neutral tensor boson $|2,0\rangle$ decayed into the two on-shell $W$-bosons in an external magnetic field when each of the $W^+$- and $W^-$-bosons have the spin projection $s_{\pm z} = 0$ is to be in the mass range

$$m_Y \geq 2m_W = 160.770 \, GeV. \qquad (40)$$

This inequality enables us to tell that when both of the $W^\pm$-bosons have the spin projection $s_{\pm z} = 0$, the mass of the single neutral tensor boson decayed into the two on-shell $W$-bosons in an external magnetic field can not be less than $2m_W = 160.770 \, GeV$.

### 2.6. Polarization state $|2, -1\rangle$

The polarization state $|2, -1\rangle$ decays according to the reactions

$$Y(J = 2, J_z = -1) = W^-(s_- = 1, s_{-z} = -1) + W^+(s_+ = 1, s_{+z} = 0), \qquad (41)$$

$$Y(J = 2, J_z = -1) = W^-(s_- = 1, s_{-z} = 0) + W^+(s_+ = 1, s_{+z} = -1). \qquad (42)$$

When both of the $W^\pm$-bosons occupy the quantum state with $n = n' = 0$ and $p_z = p_z' = 0$, the energy conservation law for the reaction channels (41) and (42) is

written as follows

$$m_Y = \sqrt{m_W^2 + eB} + \sqrt{m_W^2 + 3eB}. \tag{43}$$

When the magnetic field strength is equal to its critical value $B = B_{0W} = m_W^2/e$ and $W^-(W^+)$-boson has the spin projection $s_{-z} = -1(s_{+z} = 0)$ or $W^-(W^+)$-boson has the spin projection $s_{-z} = 0(s_{+z} = -1)$, the mass of the single neutral tensor boson decayed into the two on-shell $W$-bosons in an external magnetic field is

$$m_Y(B = B_{0W}) = (2 + \sqrt{2})m_W \cong 274.452\ GeV. \tag{44}$$

Taking into account the inequality (39) which satisfies the equation (43) we obtain the following range for the mass of the single neutral tensor boson $|2,-1\rangle$ decayed into the two on-shell $W$-bosons in an external magnetic field provided that $W^-(W^+)$-boson has the spin projection $s_{-z} = -1(s_{+z} = 0)$ or $W^-(W^+)$-boson has the spin projection $s_{-z} = 0(s_{+z} = -1)$:

$$2m_W \leq m_Y < \infty. \tag{45}$$

### 2.7. Polarization state $|2, -2\rangle$

The contribution to the polarization state $|2, -2\rangle$ comes from the reaction

$$Y(J = 2, J_z = -2) = W^-(s_- = 1, s_{-z} = -1) + W^+(s_+ = 1, s_{+z} = -1). \tag{46}$$

When both of the $W^\pm$-bosons occupy the quantum state with $n = n' = 0$ and $p_z = p'_z = 0$, the energy conservation law for the reaction channel (46) is written as follows

$$m_Y = \sqrt{m_W^2 + 3eB} + \sqrt{m_W^2 + 3eB}. \tag{47}$$

When the magnetic field strength is equal to its critical value $B = B_{0W} = m_W^2/e$ and $W^\pm$-boson has the spin projection $s_{\pm z} = -1$, the mass of the single neutral tensor boson decayed into the two on-shell $W$-bosons in an external magnetic field is

$$m_Y(B = B_{0W}) = 4m_W \cong 321.540\ GeV. \tag{48}$$

Using the inequality (39) which satisfies the equation (47) we get the following range for the mass of the single neutral tensor boson $|2,-2\rangle$ decayed into the two on-shell $W$-boson in an external magnetic field provided that $W^{\pm}$-boson has the spin projection $s_{\pm z}=-1$:

$$2m_W \leq m_Y < \infty. \tag{49}$$

## 2.8. Polarization state $|1,+1\rangle$

The polarization states (15)-(17) have the spin 1. Therefore these polarization states belong to the class of vector particles. Let us consider the decay reactions of the vector polarization states.

The polarization state $|1,+1\rangle$ decays according to the reactions

$$Y(J=1, J_z=+1) = W^-(s_-=1, s_{-z}=+1) + W^+(s_+=1, s_{+z}=0), \tag{50}$$

$$Y(J=1, J_z=+1) = W^-(s_-=1, s_{-z}=0) + W^+(s_+=1, s_{+z}=+1). \tag{51}$$

When both of the $W^{\pm}$-bosons occupy the quantum state with $n=n'=0$ and $p_z = p'_z = 0$, the corresponding energy conservation law for the reaction channels (50) and (51) is written on the basis of the expressions (4) and (5) as follows:

$$m_Y = \sqrt{m_W^2 - eB} + \sqrt{m_W^2 + eB}. \tag{52}$$

On the basis of the inequality (21) and the equality (52) we derive the following mass range

$$\sqrt{2}m_W \leq m_Y \leq 2m_W. \tag{53}$$

So, the mass of the single neutral boson with the spin $J=1$ and the spin projection $J_z = +1$ which is the vector polarization state $|1,+1\rangle$ is in the mass range (53).

## 2.9. Polarization state $|1, 0\rangle$

The contributions to the vector polarization state $|1, 0\rangle$ come from the reactions

$$Y(J=1, J_z=0) = W^-(s_-=1, s_{-z}=+1) + W^+(s_+=1, s_{+z}=-1), \tag{54}$$

$$Y(J=1, J_z=0) = W^-(s_-=1, s_{-z}=-1) + W^+(s_+=1, s_{+z}=+1). \tag{55}$$

When both of the $W^{\pm}$-bosons occupy the quantum state with $n=n'=0$

and $p_z = p'_z = 0$, the corresponding energy conservation law is written as

$$m_Y = \sqrt{m_W^2 - eB} + \sqrt{m_W^2 + 3eB} \qquad (56)$$

for the reaction channels (54) and (55). The equality (56) shows that the strength of an external magnetic field changes in the range determined with the inequality (21). Using the inequality (21) and the equality (56) we obtain the following mass range

$$2m_W \leq m_Y \leq \frac{4\sqrt{3}}{3} m_W, \qquad (57)$$

for the single neutral boson with the spin $J = 1$ and the spin projection $J_z = 0$ which is the vector polarization state $|1, 0\rangle$. When the single neutral boson with the spin $J = 1$ and the spin projection $J_z = 0$ decays into the two on-shell $W$-bosons in a constant homogenous magnetic field and one of the $W$-bosons has the spin projection $s_{\mp z} = \pm 1$, the other $W$-boson has the spin projection $s_{\pm z} = \mp 1$, the maximal value of the mass of this single neutral boson corresponds to the magnetic field strength $B = 2B_{0W}/3 = (2/3)(m_W^2/e)$.

## 2.10. Polarization state $|1, -1\rangle$

The polarization state $|1, -1\rangle$ decays according to the reactions

$$Y(J = 1, J_z = -1) = W^-(s_- = 1, s_{-z} = 0) + W^+(s_+ = 1, s_{+z} = -1), \qquad (58)$$

$$Y(J = 1, J_z = -1) = W^-(s_- = 1, s_{-z} = -1) + W^+(s_+ = 1, s_{+z} = 0). \qquad (59)$$

When both of the $W^\pm$-bosons occupy the quantum state with $n = n' = 0$ and $p_z = p'_z = 0$, the energy conservation law for the reaction channels (58) and (59) reads as follows

$$m_Y = \sqrt{m_W^2 + eB} + \sqrt{m_W^2 + 3eB}. \qquad (60)$$

When the magnetic field strength is equal to the critical value $B = B_{0W} = m_W^2/e$ and $W^-(W^+)$-boson has the spin projection $s_{-z} = 0 (s_{+z} = -1)$ or $s_{-z} = -1 (s_{+z} = 0)$, the mass of the single neutral vector boson decayed into the two on-shell $W$-bosons in an external magnetic field is $m_Y(B = B_{0W}) = (2 + \sqrt{2})m_W$. Taking into account the inequality (39) which satisfies the equation (60) we obtain the following range for the mass of the single neutral vector boson $|1,-1\rangle$ decayed into the two on-shell $W$-bosons in an external magnetic field provided that $W^-(W^+)$-boson has the spin projection $s_{-z}=0\,(s_{+z}=-1)$ or $W^-(W^+)$-boson has the spin projection $s_{-z}=-1\,(s_{+z}=0)$:

$$2m_W \leq m_Y < \infty. \tag{61}$$

## 2.11. Polarization state $|0, 0\rangle$

The polarization state (18) has the spin $0$. Therefore this polarization state belongs to the class of scalar particles. Let us consider the decay reactions of the scalar polarization state. The contributions to the polarization state $|0, 0\rangle$ are given by the reactions

$$Y(J=0, J_z=0) = W^-(s_-=1, s_{-z}=+1) + W^+(s_+=1, s_{+z}=-1), \tag{62}$$

$$Y(J=0, J_z=0) = W^-(s_-=1, s_{-z}=-1) + W^+(s_+=1, s_{+z}=+1), \tag{63}$$

$$Y(J=0, J_z=0) = W^-(s_-=1, s_{-z}=0) + W^+(s_+=1, s_{+z}=0). \tag{64}$$

When both of the $W^\pm$-bosons occupy the quantum state with $n=n'=0$ and $p_z=p'_z=0$, the corresponding energy conservation law is written as

$$m_Y = \sqrt{m_W^2 - eB} + \sqrt{m_W^2 + 3eB} \tag{65}$$

for the reaction channels (62) and (63). The equality (65) shows that the strength of an external magnetic field changes in the range determined with the inequality (21). Using the inequality (21) and the energy conservation law (65) we obtain the following mass range

$$2m_W \leq m_Y \leq \frac{4\sqrt{3}}{3} m_W \tag{66}$$

for the single neutral boson with the spin $J=0$ and the spin projection $J_z=0$ which is the scalar polarization state $|0, 0\rangle$. When the single neutral boson with the spin $J=0$ and the spin projection $J_z=0$ decays into the two on-shell $W$-bosons in a constant homogenous magnetic field, one of the $W$-bosons has the spin projection $s_{\mp z} = \pm 1$, the other $W$-boson has the spin projection $s_{\pm z} = \mp 1$.

Now let us consider the decay of a single neutral scalar boson into the two on-shell $W$-bosons in an external magnetic field when both of the $W^{\pm}$- bosons occupy the state with the spin projection $s_{\pm z} = 0$. On the basis of the energy and spin projection conservation laws we can write for the reaction channel (64) the following equality

$$m_Y = \sqrt{m_W^2 + eB} + \sqrt{m_W^2 + eB} \ . \qquad (67)$$

The minimal value of the mass of a single neutral scalar boson $|0, 0\rangle$ decayed into the two on-shell $W$-bosons according to the the channel (64) is

$$m_Y = 2m_W \cong 160.770 \ GeV \qquad (68)$$

and corresponds to the magnetic field strength of $B = 0$. At the critical value of the magnetic field strength $B = B_{0W} = m_W^2/e$ the mass of the single neutral scalar boson produced via the channel (64) in an external magnetic field is

$$m_Y(B = B_{0W}) = 2\sqrt{2}m_W \cong 227.363 \ GeV \ . \qquad (69)$$

Taking into account the inequality (21) and the energy conservation law (67) we can come to the conclusion that when the single neutral scalar boson $|0,0\rangle$ decays into the two on-shell $W$-bosons in an external magnetic field and each of the $W^+$- and $W^-$-bosons have the spin projection $s_{\pm z} = 0$, the mass of the single neutral scalar boson is to be in the range

$$m_Y \geq 2m_W = 160.770 \ GeV \ . \qquad (70)$$

This inequality enables us to tell that when both of the $W^{\pm}$-bosons have the spin projection $s_{\pm z} = 0$, the mass of the single neutral scalar boson decayed into the two on-shell $W$-bosons in an external magnetic field can not be less than $2m_W = 160.770 \ GeV$.

### 3. CHARGE CONJUGATION, SPECTRAL TERM AND INTRINSIC PARITY OF

# NEUTRAL BOSON WITH SPIN $J = 2$, SPIN PROJECTION $J_z = +2$ AND MASS AROUND 126 $GeV$

Now let us consider the charge conjugation $C$ of the indicated particle with the spin $J = 2$ and the spin projection $J_z = +2$. As we noted the new neutral boson with the mass $125.3\ GeV/126\ GeV$ has been observed in the decay channels

$$Y \rightarrow \gamma + \gamma, \tag{71}$$

$$Y \rightarrow Z + Z^*. \tag{72}$$

Both the ATLAS and CMS Collaborations also reported about the existence of some excess of the events connected with the decay $H \rightarrow WW^* \rightarrow \ell\nu\ell\nu$ [28, 29] (see also Ref. [35]). Recently the ATLAS Collaboration has reported the results on the Standard Model Higgs boson search in the $H \rightarrow WW^* \rightarrow e\nu\mu\nu$ decay mode [43]. This search has been performed using proton-proton collision data corresponding to an integrated luminosity of $5.8\ fb^{-1}$ at a centre-of-mass energy of $8\ TeV$ collected during 2012. The results of this search show that an excess of events over the expected background is observed corresponding to 3.2 standard deviations [43]. This decay channel can be written as follows

$$Y \rightarrow W^- + W^+. \tag{73}$$

Now we consider the diphoton decay channel (71). The eigenvalue of the charge conjugation for the photon is $C_\gamma = -1$ [50]. A charge conjugation is also a multiplicative quantum number. Using this fact and the diphoton decay reaction (71) we can write

$$C \equiv C_Y = C_\gamma C_\gamma = (-1)^2 = +1 \tag{74}$$

for the charge conjugation of the observed $Y$- boson. On the other hand there is the following relation between the charge conjugation $C$, the orbital quantum number $L$ and the spin $J$

$$C = (-1)^{L+J}. \tag{75}$$

Using the relations (74), (75) and taking into account the spin $J = 2$ of the new neutral boson $Y$ we derive that the value of $L$ can be only even numbers and the minimal value of the orbital quantum number is $L = 0$ that corresponds to the $S$ spectral term. The single neutral boson with the mass around $126\,GeV$ can be in principle considered as both a fundamental particle and a composite particle like a bound state. For generality, here we do not distinguish these two indicated cases. To determine all the characteristics of the spectral term of the single neutral boson with the mass around $126\,GeV$ we use the standard spectral term notation $(n_r + 1)^{2J+1}L_M$ where $M = L+J, L+J-1, ..., |L-J|$ corresponds to the total angular momentum, $n_r$ is the "radial" quantum excitation number. $n_r = 0$ corresponds to the lowest state in the spectrum. So, taking into account that $J = 2$, $L = 0$ we have $M = J = 2$ and obtain the spectral term $1^5 S_2$ for the neutral boson with the mass around $126\,GeV$, the spin $J = 2$ and the spin projection $J_z = +2$. And now let us determine the $P$-parity of the indicated particle with the spin $J = 2$ and the spin projection $J_z = +2$. The new neutral particle with the mass $125.3\,GeV/126\,GeV$ is a boson. Therefore the wave function of this particle is to be an even function

$$\psi(r, \vartheta, \varphi) \to \psi(-r, \vartheta - \pi, \varphi + \pi) = \psi(r, \vartheta, \varphi). \tag{76}$$

On the other hand according to the spatial inversion of the coordinates in the spherical system

$$\psi(r, \vartheta, \varphi) \to \psi(-r, \vartheta - \pi, \varphi + \pi) = I\psi(r, \vartheta, \varphi) = P(-1)^L \psi(r, \vartheta, \varphi), \tag{77}$$

where $P$ is the intrinsic parity of the neutral boson with the mass $125.3\,GeV/126\,GeV$ and $I$ is the total parity of that particle. On the basis of (76) and (77) the following result

$$I = (-1)^L P = 1 \tag{78}$$

is derived for the total parity $I$. $L = 0$ value of the orbital quantum number and the relation (78) enables us to determine the intrinsic parity of the neutral boson $Y$ discovered in the CMS and ATLAS experiments at the LHC

$$P \equiv P_Y = I = +1. \tag{79}$$

On the basis of the equalities (74) and (79) we determine that

$$CP \equiv C_Y P_Y = +1 \tag{80}$$

So, the neutral boson with the mass 125.3 *GeV* /126 *GeV* discovered in the CMS and ATLAS experiments at the LHC is a new neutral boson with the spin $J = 2$ (a tensor particle) and the spin projection $J_z = +2$ that is not included in the Standard Model. Both the $P$- parity and the charge conjugation $C$ of this new particle are $+1$, i.e. this is a new neutral boson with the quantum numbers $J^{PC} = 2^{++}$, the spin projection $J_z = +2$ and $CP = +1$.

## 4. DETERMINATION OF WEAK ISOSPIN $T$, ITS THIRD COMPONENT $T_z$ AND WEAK HYPERCHARGE $Y_W$ OF NEUTRAL BOSON WITH MASS AROUND 126 *GeV*

The $W^+ (W^-)$-boson is characterized with the weak isospin $T = 1$ and its positive (negative) projection onto the $Oz$-axis $T_z = +1$ ($T_z = -1$). We denote the $W^+$-boson state as

$$\left|W^+(T=1, T_z=+1)\right\rangle = |1, +1\rangle \quad (81)$$

and $W^-$-boson state as

$$\left|W^-(T=1, T_z=-1)\right\rangle = |1, -1\rangle. \quad (82)$$

Using the Clebsh-Gordan coefficients [41, 49] and the fact that $W^\pm$-bosons can not be in the isospin state with

$$\left|W^\pm(T=1, T_z=0)\right\rangle = |1, 0\rangle \quad (83)$$

we obtain the following relation for the addition of the weak isospins of the $W^+ (W^-)$-bosons for the reaction (1):

$$|1, -1; 1+1\rangle = \frac{1}{\sqrt{6}}|2, 0\rangle - \frac{1}{\sqrt{2}}|1, 0\rangle + \frac{1}{\sqrt{3}}|0, 0\rangle. \quad (84)$$

From the last expression we see that the produced neutral boson $Y$ with the mass 125.3 *GeV* /126 *GeV* can be in three weak isospin states

$$|2, 0\rangle = \frac{1}{\sqrt{6}}|1, -1; 1, +1\rangle, \quad (85)$$

$$|1,0\rangle = -\frac{1}{\sqrt{2}}|1,-1;1,+1\rangle, \tag{86}$$

$$|0,0\rangle = \frac{1}{\sqrt{3}}|1,-1;1,+1\rangle. \tag{87}$$

with different coefficients (weights). In all three states the third component of the weak isospin of the produced neutral boson $Y$ with the mass 125.3 $GeV$ / 126 $GeV$ is $T_z = 0$. It is the result of the conservation law of the third component of the weak isospin. By the way, the weak isospin conservation law namely relates the conservation of $T_z$.

It should be noted that the electric charge $Q$ is related to the third component $T_z$ of the weak isospin and the weak hypercharge $Y_W$ by the relation

$$Q = T_z + \frac{Y_W}{2}. \tag{88}$$

Taking into account that the neutral boson $Y$ with the mass 125.3 $GeV$ / 126 $GeV$ is a neutral particle $(Q = 0)$ and the third component of the weak isospin of this boson is zero $(T_z = 0)$ we obtain

$$Y_W = 0. \tag{89}$$

The weak isospin determines the number of particles or energy states in the given isomultiplet. The weak isospin $T = 2$ corresponds to five particles or energy states. These five particles or energy states correspond to the polarization states (10)-(14) of the neutral boson with the spin $J = 2$. The weak isospin $T = 1$ corresponds to the three polarization states (15)-(17) of a neutral boson with the spin $J = 1$. The weak isospin $T = 0$ corresponds to the neutral boson with the spin $J = 0$.

### 5. PRODUCTION OF SINGLE NEUTRAL BOSON FROM VACUUM IN STRONG EXTERNAL ELECTRIC FIELD

An electromagnetic vacuum is unstable in an electric field of the strength

$$B_{0e} = E_{0e} = \frac{m_e^2 c^3}{e\hbar} \tag{90}$$

or

$$B_{0e} = E_{0e} = \frac{m_e^2}{e} \qquad (91)$$

if we have dealing with the system of units where the Planck constant $\hbar$ and the light speed in a vacuum $c$ are equal to unit, i. e. $\hbar = c = 1$. Since $W^{\pm}$-bosons are charged particles we can consider an electroweak vacuum as "a sea" filled with virtual $W^+W^-$-pairs. Of course, the vacuum is also filled with quark and antiquark pairs, charged lepton and antilepton pairs. Here we only consider the vacuum consisting of virtual $W^+W^-$-pairs. These $W^{\pm}$-bosons are off-shell. When virtual (off-shell) $W^{\pm}$-bosons existing in the vacuum gain the energy $2m_W c^2$ in an electric field of the strength $E_{0W} = B_{0W}$ in the distance equal to the Compton wave length of a $W^{\pm}$-boson $\hbar/m_W c$, $W^{\pm}$-boson pairs are proued from the vacuum, i. e.

$$2eE_{0W} \frac{\hbar}{m_W c} = 2m_W c^2 \qquad (92)$$

or

$$B_{0W} = E_{0W} = \frac{m_W^2 c^3}{e\hbar} \left( = \frac{m_W^2}{e} \right). \qquad (93)$$

And now let us consider the production of a single neutral particle $Y$ from the vacuum at the expense of a $W^{\pm}$-pair in an external electric field. When virtual $W^{\pm}$-bosons existing in the vacuum gain the energy $m_Y c^2$ in an electric field of the strength $E_{0WY} = B_{0WY}$ in the distance equal to the Compton wave length of a $W^{\pm}$-boson, a single neutral particle $Y$ is produced from the vacuum, i.e.

$$2eE_{0WY} \frac{\hbar}{m_W c} = m_Y c^2 \qquad (94)$$

or

$$B_{0WY} = E_{0WY} = \frac{m_Y m_W c^3}{2e\hbar} \left( = \frac{m_Y m_W}{2e} \right). \qquad (95)$$

Now let us apply the Gauss theorem to the case when the strength of an electric field is $E_{0W} = \dfrac{m_W^2}{e}$ and the charge contained within the sphere of a radius $r_W$ is $e$

$$\dfrac{m_W^2}{e} \times 4\pi r_W^2 = e. \tag{96}$$

The radiative corrections to the electric charge at small distances will be taken a bit later. We find from the last expression the radius $r_W$

$$r_W = \dfrac{\sqrt{\alpha}}{m_W} \tag{97}$$

where $\alpha = e^2/4\pi$ is the fine structure constant.

Now we apply the Gauss theorem to the case when the strength of an electric field is $E_{0WY} = m_Y m_W / 2e$ and the charge $q_Y$ contained within the sphere of a radius

$$r_Y = \dfrac{\sqrt{\alpha}}{m_W}\left(\dfrac{m_W}{m_Z}\right)^k \tag{98}$$

is equal to

$$q_Y = e\left(\dfrac{m_W}{m_Z}\right)^{3k}. \tag{99}$$

If we take into account the expressions (95), (98) and (99) in the Gauss theorem

$$E_{0WY} \times 4\pi r_Y^2 = q_Y, \tag{100}$$

we obtain the following result

$$m_Y = 2m_W\left(\dfrac{m_W}{m_Z}\right)^k = 2m_W \cos^k \theta_W, \tag{101}$$

where $k$ is an arbitrary real number. If we use the PDG data for the masses of a $W^\pm$-boson and $Z$-boson ($m_W \cong 80.385\ GeV$; $m_Z \cong 91.1876\ GeV$) [49], we can present some of the masses corresponding to the numbers $k = 0, \pm 1, \pm 2$.

$$m_Y = \dfrac{2m_W^3}{m_Z^2} = 124.935\,GeV,\ k = 2, \tag{102}$$

$$m_Y = \dfrac{2m_W^2}{m_Z} = 141.724\,GeV,\ k = 1, \tag{103}$$

$$m_Y = 2m_W = 160.770\, GeV, \quad k = 0, \tag{104}$$

$$m_Y = 2m_Z = 182.375\, GeV, \quad k = -1, \tag{105}$$

$$m_Y = \frac{2m_Z^2}{m_W} = 206.884\, GeV, \quad k = -2, \tag{106}$$

It is visible from the expression (102) that a single neutral particle with the mass $\approx 125\, GeV$ (the radiative corrections have not taken into account yet) can be produced from the vacuum in an electric field at the expense of the virtual (off-shell) $W^{\pm}$-boson pair fusion.

The formulae (101)-(106) show that single neutral particles with various masses can be produced from the vacuum in an external electric field at the expense of $W$-boson pair fusion.

Using (95) and (102) we can estimate the strength of an external electric field that is sufficient for production of a single neutral boson with the mass $m_Y = 2m_W^3/m_Z^2$

$$E_{0WY} = \frac{m_W^4}{em_Z^2} = \frac{m_W^2}{e}\left(\frac{m_W}{m_Z}\right)^2 \cong 0.777\frac{m_W^2}{e}. \tag{107}$$

## 6. DISCUSSION OF THE RESULTS

In this work we have investigated the single neutral bosons with spin 0, 1 and 2 decayed via the $W^-W^+$-channel and discussed the questions connected with the search of the Standard Model scalar Higgs boson in the CMS and ATLAS experiments at the Large Hadron Collider.

The obtained here results and calculations are true for both the $W^-W^+ \to Y$ reaction and $Y \to W^-W^+$ reaction. Therefore we investigated one of these reactions in a constant homogenous magnetic field. On the other hand the production of single neutral bosons with the spins 0 and 2 at the expense of the reaction $W^-W^+ \to Y$ in an external magnetic field was studied by us in [42] where we determined the spin ($J = 2$) and the spin projection ($J_z = +2$) of the single neutral boson with the mass around $126\, GeV$.

The calculations and analyses performed by us show that the spin additional rule gives nine possible polarization states. Only three of them, i.e. the polarization states $|1,1\rangle$, $|2,1\rangle$ and $|2,+2\rangle$ can, in principle, correspond to the neutral boson with the mass around 126 $GeV$. Because the calculations give the range $\sqrt{2}m_W \leq m_Y \leq 2m_W$ for the polarization states $|1,1\rangle$ and $|2,1\rangle$ and the range $0 \leq m_Y \leq 2m_W$ for the polarization state $|2,+2\rangle$. As we mentioned the observed single neutral particle with the mass around 126 $GeV$ also decays into the two photons. However, according to the Landau-Yang theorem the diphoton decay of a single neutral boson with the spin 1 is forbidden. Thus, the polarization state $|1,1\rangle$ can not be a single neutral boson with the mass around 126 $GeV$. Using the Clebsh-Gordan coefficients [41, 49] we obtain the following polarization states of the single neutral boson that can, in principle, decay into the two photons

$$|2,+2\rangle = |1,+1;1,+1\rangle, \qquad (108)$$

$$|2,0\rangle = \frac{1}{\sqrt{6}}\left(|1,+1;1,-1\rangle + |1,-1;1,+1\rangle\right), \qquad (109)$$

$$|2,-2\rangle = |1,-1;1,-1\rangle, \qquad (110)$$

$$|1,0\rangle = \frac{1}{\sqrt{2}}\left(|1,+1;1,-1\rangle - |1,-1;1,+1\rangle\right), \qquad (111)$$

$$|0,0\rangle = \frac{1}{\sqrt{3}}\left(|1,+1;1,-1\rangle + |1,-1;1,+1\rangle\right). \qquad (112)$$

where $|1,+1\rangle$ is the photon state with the spin polarization $+1$ and $|1,-1\rangle$ is the photon state with the spin polarization $-1$. As we indicated the diphoton decay is impossible for the spin 1 particle. Therefore, the polarization state $|1,0\rangle$ can not be a single neutral particle with the mass around 126 $GeV$. For the states $|2,0\rangle$, $|2,-2\rangle$ and $|0,0\rangle$ the mass of the single neutral boson satisfies the inequality $m_Y \geq 2m_W$. It means that the states $|2,0\rangle$, $|2,-2\rangle$ and $|0,0\rangle$ can not be a single neutral particle with the mass around 126 $GeV$. So, the only one remaining polarization state $|2,+2\rangle$ which corresponds to the mass range $0 \leq m_Y \leq 2m_W$ can decay into the two photons. So, the single neutral boson with the mass around $126\,GeV$ discovered by the CMS and ATLAS Collaborations corresponds to the mass range $0 \leq m_Y \leq 2m_W$. This range, in its turn, corresponds to the particle with the spin $J = 2$ and the spin projection $J_z = +2$. Thus, we come to the conclusion that the new neutral boson with the mass $125.3\,GeV/126\,GeV$ discovered in the CMS and ATLAS experiments is a new boson with the spin $J = 2$ and the spin projection $J_z = +2$.

The results obtained by us explain that why the decay modes $b\bar{b}$, $\tau\bar{\tau}$ are not observed. Because the observed neutral boson with the mass $125.3\,GeV/126\,GeV$ is not a scalar boson and these decay modes are not allowed by the spin projection conservation law (and by the spin conservation law). The particle with the spin 2 can not decay into the two fermions that have a half spin. The neutral boson with the spin 2 can decay into the two photons, $Z$- and $W^{\pm}$- bosons. These decays are allowed by the spin conservation law. Therefore the observation of $H \to Z\gamma$, $H \to ZZ^* \to 4l$ decay modes are normal. This is one of the real reasons of the observed significant deviations from the Standard Model expectation including the deficits in the $b\bar{b}$ and $\tau^+\tau^-$ channels.

The calculations and analysis show that the Standard Model Higgs boson or any other scalar neutral boson with the mass $125.3\,GeV/126\,GeV$ can not decay into the two $W$-bosons in the mass ranges $116.6\,GeV - 119.4\,GeV$ and $122.1\,GeV - 127\,GeV$ if it really exists. We are to search for a single neutral scalar boson or the Standard Model Higgs boson in the mass range above $2m_W$ if it really exists and decays into two $W$-bosons.

The calculations and analysis show that the third component of the weak isospin of the discovered neutral boson $Y$ with the mass $125.3\,GeV/126\,GeV$ is $T_z = 0$. The observed neutral boson with the mass $125.3\,GeV/126\,GeV$ corresponds to the weak isospin state $|Y(T = 2, T_z = 0)\rangle = |2, 0\rangle$. The weak isospin state $|Y(T=1, T_z=0)\rangle = |1, 0\rangle$ corresponds to the other theoretically possible neutral boson states with the spin 1. The weak isospin state $|Y(T=0, T_z=0)\rangle = |0, 0\rangle$ corresponds to the another theoretically possible neutral boson state with the spin 0.

The calculation of the third component of the weak isospin of the neutral boson $Y$ with the mass $125.3\,GeV/126\,GeV$ enables us also to determine the weak hypercharge of the observed unknown neutral boson which is $Y_W = 0$.

The mass $m_Y = 2m_W^3/m_Z^2 \cong 125\,GeV$ determined theoretically by us corresponds to the neutral boson with the mass $125.3\,GeV/126\,GeV$. To determine the mass by the used theoretical methods more precisely the radiative corrections should be taken into account. As we indicated the particle with the spin 2 can be in five possible polarization states. The polarization states $|2, +2\rangle$ and $|2, 1\rangle$ only satisfies the condition of a bound state. The polarization state $|2, +2\rangle$ that corresponds, in our opinion, to the single neutral boson with the mass around $126\,GeV$ can be considered as a bound state [51]. On the other hand it is also possible to consider the state $|2, +2\rangle$ as a fundamental particle that is a carrier (quantum) of interactions between bosons, at least between electroweak bosons. To clarify this situation more and new experimental data are required.

If the experimental data proves that the observed neutral boson with the mass around $126\,GeV$ is a particle with the spin 2, then our thinking on the space-time and on its dimensions [52, 53] will change essentially. The interest to the investigations connected with the spin 2 particles increases day after day [54]. The existence of the particle with the spin 2 would indicate that the world we live has additional dimensions besides known four dimensions.

## 7. CONCLUSIONS

We discussed the questions connected with the search of the Standard Model scalar Higgs boson in the CMS and ATLAS experiments at the Large Hadron Collider. It is shown that the Higgs boson or any other scalar boson with the mass $125.3\,GeV/126\,GeV$ can not decay into the two on-shell $W$-bosons in the mass ranges $116.6\,GeV - 119.4\,GeV$ and $122.1\,GeV - 127\,GeV$ in an external magnetic field if it really exists. The impossibility of decay of a single neutral scalar boson into the two on-shell $W$-bosons in the mass range below $160.770\,GeV$ because of the energy and spin projection conservation laws and the possible decay of a single neutral boson with the spin $J = 2$ and the spin projection $J_z = +2$ into the two on shell $W$-bosons in an external constant homogenous magnetic field in that region enable us to come to the conclusion that the single neutral boson with the mass $125.3\,GeV/126\,GeV$ discovered in the CMS and ATLAS experiments is neither the Standard Model Higgs boson nor a scalar boson at all. The neutral boson with the mass $125.3\,GeV/126\,GeV$ discovered in the CMS and ATLAS experiments is a new neutral boson with the spin $J = 2$ (a tensor particle) and the spin projection $J_z = +2$ that is not included in the Standard Model. Both the intrinsic parity $P$ and charge conjugation $C$ of this new particle are $+1$. So, the newly discovered particle with the mass $125.3\,GeV/126\,GeV$ is a neutral boson with the spin $J = 2$ (a tensor particle) and the spin projection $J_z = +2$ that is characterized by $J^{PC} = 2^{++}$ quantum numbers under the intrinsic parity $P$ and the charge conjugation $C$ and $CP = +1$. The other quantum characteristics of this neutral boson are as follows: the orbital quantum number $L = 0$, the weak isospin $T = 2$, the third component of the weak isospin $T_z = 0$ and the weak hypercharge $Y_W = 0$. The quantum state of the discovered neutral boson is described by the $1^5S_2$ spectral term. At the same time the observed neutral boson with the mass around $126\,GeV$ can be considered as a fundamental particle that is a carrier (quantum) of the interactions between bosons, at least between electroweak bosons. To clarify this situation more and new experimental data are required.

**REFERENCES**

1. S. L. Glashow, Nucl. Phys. **22**, 579 (1961).
2. S. Weinberg, Phys. Rev. Lett. **19**, 1264 (1967).
3. A. Salam, in *Elementary Particle Theory*, edited by N. Svartholm (Wiley, New

York 1968; Almqvist and Wiksell, Stockholm 1968).

4. M. Gell-Mann, Phys. Rev. Lett. **8**, 214 (1964).
5. H. Fritzsch, M. Gell-Mann, and H. Leutwyler, Phys. Lett. B **47**, 365 (1973).
6. D. I. Gross, F. Wilczek, Phys. Rev. Lett. **30**, 1343 (1973).
7. H. D. Politzer, Phys. Rev. Lett. **30**, 1346 (1973).
8. G. Arnison *et al.*, Phys. Lett. B **122**, 103 (1983).
9. G. Arnison *et al.*, Phys. Lett. B, **126,** 398 (1983).
10. G. Arnison *et al.*, Phys. Lett. B, **129,** 273 (1983).
11. M. Banner *et al.*, Phys. Lett. B, **122,** 476 (1983).
12. P. Bagnia *et al.*, Phys. Lett. B, **129,** 130 (1983).
13. F. Englert and R. Brout, Phys. Rev. Lett. **13,** 321 (1964).
14. P. W. Higgs, Phys. Lett. **12**, 132 (1964).
15. P. W. Higgs, Phys. Rev. Lett. **13,** 508 (1964).
16. G. S. Guralnik, C. R. Hagen, and T. W. B. Kibble, Phys. Rev. Lett. **13**, 585 (1964)
17. P. W. Higgs, Phys. Rev. **145**, 1156 (1966).
18. T. W. B. Kibble, Phys. Rev. **155**, 1554 (1967).
19. J. Goldstone, Nuovo Cim. **19**, 154 (1961).
20. J. Goldstone, A. Salam, and S. Weinberg, Phys. Rev. **127**, 965 (1962).
21. The ALEPH, CDF, D0, DELPHI, L3, OPAL, SLD Collaborations, the LEP Electroweak Working Group, the Tevatron Electroweak Working Group, the SLD Electroweak and Heavy Flavour Groups, CERN-PH-EP-2010-095, arXiv: 1012.2367 [hep-ex], 2011.
22. The ALEPH, DELPHI, L3, OPAL, SLD Collaborations, the LEP Electroweak Working Group, the SLD Electroweak and Heavy Flavour Groups, Phys. Rept. **427,** 257 (2006).
23. LEP Working Group for Higgs boson searches, ALEPH, DELPHI, L3 and OPAL Collaborations, Phys. Lett. B **565,** 61 (2003), arXiv: 0306033 [hep-ex], 2003.
24. The CDF and D0 Collaborations, and the Tevatron New Phenomena and Higgs Working Group, Combined CDF and D0 Search for Standard Model Higgs Boson

Production with up to 10.0 fb$^{-1}$ of Data, arXiv: 1203.3774 [hep-ex], 2012.

25. T. Aaltonen *et al.* (CDF Collaboration), Phys. Rev. Lett. **109**, 111802 (2012), arXiv: 1207.1707 [hep-ex], 2012.

26. V. M. Abazov, *et al.* (D0 Collaboration), Phys. Rev. Lett. **109**, 121802 (2012), arXiv: 1207.6631 [hep-ex], 2012.

27. T. Aaltonen *et al.*, (CDF and D0 Collabarations), Phys. Rev. Lett. **109**, 071804 (2012), arXiv: 1207.6436 [hep-ex], 2012.

28. G. Aad *et al.* (ATLAS Collaboration), Phys. Lett. B **716**, 1 (2012).

29. S. Chatrchyan *et al.* (CMS Collaboration), Phys. Lett. B **716**, 30 (2012).

30. G. Aad *et al.* ATLAS Collaboration, Phys. Lett. B **710**, 49 (2012), arXiv: 1202.1408 [hep-ex], 2012.

31. S. Chatrchyan *et al.* (CMS Collaboration), Phys. Lett. B **710**, 26 (2012), arXiv: 1202.1488 [hep-ex], 2012.

32. G. Aad *et al.* (ATLAS Collaboration), Phys. Rev. D **86**, 032003 (2012), arXiv: 1207.0319 [hep-ex], 2012.

33. A. Djouadi, Phys. Rept. **457**, 1 (2008), arXiv: 0503172 [hep-ph], 2005.

34. A. Djouadi, Pramana **79**, 513 (2012), arXiv:1203.4199 [hep-ph], 2012.

35. V. Rubakov, Uspekhi fizicheskikh nauk, **182 (10)**, 1017 (2012) [Phys. Usp. **55 (10)**, 2012]

36. L. B. Okun, *Leptons and Quarks* (North-Holland, Amsterdam, 1984).

37. A. I. Vainshtein, M. B. Voloshin, V. I. Zakharov, and M. A. Shifman, Yad. Fiz. **30**, 1368 (1979) [Sov. J. Nucl. Phys. **30**, 711 (1979)].

38. J. R. Ellis, M. K. Gaillard, and D. V. Nanopoulos, Nucl. Phys. B **106**, 292 (1976).

39. L. D. Landau, Dokl. Akad. Nauk USSR **60**, 207 (1948).

40. C. N. Yang, Phys. Rev. **77**, 242 (1950).

41. G. Källén, *Elementary Particle Physics* (Addison-Wesley, Reading, MA, 1964).

42. V. A. Huseynov, arXiv:1212.5830 [physics.gen-ph].

43. The ATLAS Collaboration. ATLAS-CONF-2012-098; ATLAS-COM-CONF-2012-138.

44. T. Goldman, W. Tsai, and A. Yildiz, Phys. Rev. D **8**, 1926 (1972).

45. A. Salam, J. Strathdee, Nucl. Phys. B **90**, 203 (1975).

46. V. V. Skalozub, Yad. Fiz. **37**, 474 (1983).

47. V. V. Skalozub, Yad. Fiz. 4**3**, 1045 (1986) [Sov. J. Nucl. Phys, **43**, 665 (1986)].

48. D. Grasso, H. R. Rubinstein, Phys. Rept. **348**, 163 (2001).

49. J. Beringer *et al*., Particle Data Group, PR D **86**, 010001 (2012).

50. D. McMahon, *Quantum Field Theory Demystified* (McGraw-Hill, New York, 2008).

51. A. B. Arbuzov, I. V. Zaitsev, Phys. Rev. D **85**, 093001 (2012).

52. N. Arkani-Hamed, S. Dimopoulos and G. R. Dvali, Phys. Lett. B **429**, 263 (1998).

53. N. Arkani-Hamed, S. Dimopoulos and G. R. Dvali, Phys. Rev. D **59**, 086004 (1999).

54. Ch.-Q. Geng, D. Huang, Y. Tang and Y.-L. Wu, arXiv: 1210.5103[hep-ph].